\documentclass[a4paper,12pt]{article}

\usepackage{enumerate}
\usepackage{graphicx}
\usepackage{amsmath}
\usepackage{amssymb}
\usepackage{hyperref}
\usepackage{tabularx}
\newcolumntype{M}{>{$}c<{$}}
\usepackage[matrix,arrow,curve]{xy}

\numberwithin{equation}{section} \numberwithin{figure}{section}
\numberwithin{table}{section}

\def\papertitlepage{\baselineskip 3.5ex\thispagestyle{empty}}
\def\Title#1{\baselineskip 1cm \vspace{1.5cm}%
  \begin{center}{\Large\bf #1}\end{center}\vspace{0.5cm}}
\def\Authors#1{\begin{center}\renewcommand{\thefootnote}{\fnsymbol{footnote}}{\it #1}\end{center}}
\def\Abstract{\vspace{1.0cm}%
  \begin{center}{\large\bf Abstract}\end{center}}

\renewenvironment{thebibliography}{\pagebreak[3]\par\vspace{0.6em}
\begin{flushleft}{\large \bf References}\end{flushleft}
\vspace{-1.0em}

\begin{enumerate}\if@twocolumn\baselineskip=0.6em\itemsep -0.2em
\else\itemsep -0.2em\fi\labelsep 0.1em}{\end{enumerate} }

\begin{document}
{\papertitlepage \vspace*{0cm} {\hfill
\begin{minipage}{4.2cm}
CCNH-UFABC 2014\par\noindent April, 2014
\end{minipage}}
\Title{Comments on multibrane solutions in cubic superstring field
theory}
\Authors{{\sc E.~Aldo~Arroyo${}$\footnote{\tt
aldo.arroyo@ufabc.edu.br}}
\\
Centro de Ci\^{e}ncias Naturais e Humanas, Universidade Federal do ABC \\[-2ex]
Santo Andr\'{e}, 09210-170 S\~{a}o Paulo, SP, Brazil ${}$ }
} 

\vskip-\baselineskip
{\baselineskip .5cm \Abstract In a previous paper, we have studied
multi-brane solutions in the context of cubic superstring field
theory. The kinetic term of the action was computed for these
multi-brane solutions, and for the evaluation of the energy, the
equation of motion contracted with the solutions itself was simply
assumed to be satisfied. In this paper, we compute the cubic term
of the action and discuss the validity of the previous assumption.
Additionally, we evaluate the Ellwood's gauge invariant
observable.

 }
\newpage
\setcounter{footnote}{0}
\tableofcontents

\section{Introduction}
Schnabl's work on the first analytic solution in open bosonic
string field theory \cite{Schnabl:2005gv} can be considered the
first step towards the analytic understanding of string field
theory. After the publication of Schnabl's seminal paper, a
remarkable amount of work has been done concerning the analysis of
the tachyon vacuum solution and the construction of associated
solutions by algebraic techniques
\cite{Okawa:2006vm,Fuchs:2006hw,Rastelli:2006ap,Okawa:2006sn,Erler:2012qr,Ellwood:2009zf,
Schnabl:2010tb,Erler:2009uj,AldoArroyo:2009hf,Arroyo:2009ec,
Arroyo:2011zt,Bonora:2013cya,Masuda:2012kt}. For instance, the
tachyon vacuum solution was rewritten in terms of basic string
fields constructed out of elements in the $KBc$ subalgebra
\cite{Erler:2006hw,Erler:2006ww,Zeze:2011zz,Zeze:2010jv,Arroyo:2010sy,AldoArroyo:2011gx}.
Using the elements of this subalgebra, Murata and Schnabl have
constructed a family of solutions known as the multi-brane
solutions \cite{Murata:2011ex}. Depending on the analytic
properties of a function which parameterizes the solutions, it has
been shown that the evaluation of the energy leads to an answer
compatible with solutions that describe multiple coincident
D-branes.

Although various calculations associated with the multi-brane
solutions, such as the evaluation of gauge-invariant observables,
provide expected results, there are subtleties involved in the
computations. Since the solutions can have expressions which are
either divergent or anomalous, they must be treated with due care.
In a recent set of papers
\cite{Murata:2011ep,Hata:2011ke,Hata:2012cy}, the authors have
analyzed the existence of possible anomalies in the evaluation of
the gauge-invariant observables. The origin of
these anomalies are related to the violation of some regularity
conditions imposed on the function that parameterizes the
solutions \cite{Murata:2011ep}. As Murata and Schnabl have pointed
out the status of the multi-brane solutions might be analogous to
the tachyon vacuum solution without the phantom term.

The construction of analytic solutions in the modified cubic
superstring field theory \cite{Arefeva:1989cp} naively follows the
prescriptions used in the bosonic case. Since these two theories
have a similar cubic-like interaction term and the string field
products are based on Witten's associative star product
\cite{Witten:1985cc}, the bosonic results admit quite
straightforward extensions to the superstring case
\cite{Erler:2007xt,Kroyter:2009bg,Aref'eva:2008ad,Gorbachev:2010zz,Erler:2010pr}.
For instance, the $KBc$ subalgebra can be extended to the
$KBc\gamma$ subalgebra which includes the superstring ghost field
$\gamma$
\cite{Aref'eva:2009sj,Arefeva:2010yd,Arroyo:2010fq,Inatomi:2011an}.
Using this subalgebra, we have studied the multi-brane solutions
in the context of the modified cubic superstring field theory
\cite{AldoArroyo:2012if}.

As in the bosonic case, by evaluating the energy associated to
these solutions, we have shown that the solutions can be
interpreted as describing multiple coincident D-branes.
Nevertheless, for the evaluation of the energy, the equation of
motion contracted with the solutions itself was simply assumed to
be satisfied. In this paper, we compute the cubic term of the
action and discuss the validity of the previous assumption.
Additionally, we evaluate the Ellwood's gauge invariant observable
\cite{Ellwood:2008jh} for the multi-brane solutions. It turns out
that the energy computed from the action and from the Ellwood's
invariant will agree provided that the function that parameterizes
the multi-brane solutions satisfies appropriate holomorphicity
conditions that are similar to the bosonic case
\cite{Murata:2011ep}.

The paper is organized as follows. In section 2, we review the multi-brane solutions in the
modified cubic superstring field theory. In section 3, we compute the cubic term of the
action for the multi-brane solutions. In section 4, the Ellwood's
gauge invariant overlap for the multi-brane solutions will be
evaluated. In section 5, a summary and further directions of
exploration are given.

\section{Review of the multi-brane solutions in the cubic superstring
field theory}

In this section, a short review of the multi-brane solutions in
the modified cubic superstring field theory will be given. In our
previous paper \cite{AldoArroyo:2012if}, using the prescription
developed in reference \cite{Arroyo:2010fq}, we have derived the
multi-brane solutions by performing a gauge transformation over an
identity based solutions. Here instead of employing that
prescription, we will adopt the standard procedure, namely we are
going to write the solutions as a pure gauge form. It turns out
that solutions given in this way naively satisfy the string field
equation of motion \cite{Okawa:2006vm}.

Since the algebraic structure of the modified cubic superstring
field theory is similar to the open bosonic string field theory,
the bosonic results admit quite straightforward extensions to the
superstring case. For instance, the $KBc$ subalgebra of the
bosonic string field theory can be extended to the $KBc\gamma$
subalgebra which includes the superstring ghost field $\gamma$
\cite{Arroyo:2010sy,Erler:2007xt,Gorbachev:2010zz,Arroyo:2010fq}.

Employing the elements of the $KBc\gamma$ subalgebra, we construct
a rather generic solution which can be written as a pure gauge
form $\Psi=U Q U^{-1}$ with the string field $U$ defined by
\begin{eqnarray}
\label{sec207} U = 1-F Bc F \;\; , \;\;\;\; U^{-1} = 1+
\frac{F}{1-F^2} Bc F,
\end{eqnarray}
where $F$ is a function of $K$, and $B$, $c$ are the
elements of the $KBc\gamma$ subalgebra. These basic string fields
satisfy the usual algebraic relations
\begin{align}
&\{B,c\}=1\, , \;\;\;\;\;\;\; [B,K]=0 \, , \;\;\;\;\;\;\;
B^2=c^2=0
\, , \nonumber\\
\label{sec208} \partial c = [K&,c] \, , \;\;\;\;\;\;\;
\partial \gamma  = [K,\gamma] \, , \;\;\;\;\;\;\; [c,\gamma]=0 \, ,
\;\;\;\;\;\;\; [B,\gamma]=0 \, ,
\end{align}
and have the following BRST variations
\begin{eqnarray}
\label{sec209} QK=0 \, , \;\;\;\;\;\; QB=K \, , \;\;\;\;\;\;
Qc=cKc-\gamma^2 \, , \;\;\;\;\;\; Q\gamma=c \partial \gamma
-\frac{1}{2} \gamma
\partial c \, .
\end{eqnarray}

Performing some algebraic manipulations with these basic string
fields, and using equations (\ref{sec207})-(\ref{sec209}) we can
write the following solution
\begin{eqnarray}
\label{solpsi2} \Psi= F c \frac{KB}{1-F^2} c F + F B
\gamma^2 F,
\end{eqnarray}
which formally satisfies the string field equation of motion $Q
\Psi + \Psi \Psi =0$, where $Q$ is the BRST operator of the open
Neveu-Schwarz superstring theory. Since the solution for the
superstring case (\ref{solpsi2}) is almost
similar to the bosonic solution $\Psi_{\text{bos}}= F c \frac{KB}{1-F^2} c F$,
the second term on the right-hand side of equation (\ref{solpsi2}) is commonly known as
the \textit{superstring correction} \cite{Erler:2007xt}.

In the framework of the modified cubic superstring field theory,
the solution (\ref{solpsi2}) has been studied for the specific
cases: $F^2=e^{-K}$ and $F^2=1/(1+K)$, where it was shown that the
solution characterizes the tachyon vacuum solution
\cite{Erler:2007xt,Gorbachev:2010zz}. It is interesting to note
that, as argued in reference \cite{Erler:2007xt}, from an analytic
perspective the proposed tachyon vacuum solution in the modified
cubic superstring field theory appears to be as regular as
Schnabl's original solution for the bosonic string. Nevertheless,
from the perspective of the level expansion the situation is
unclear, though to be honest, the analysis of the energy for the
tachyon vacuum solution using the usual Virasoro $L_0$ level
expansion has not yet been carried out. Relevant considerations
related to the gauge equivalence of the tachyon vacuum solutions
were properly analyzed in reference \cite{Arefeva:2010yd}.

The evaluation of the energy for a class of analytic solutions of
the form (\ref{solpsi2}) for a generic function $F(K)$ was
performed in reference \cite{AldoArroyo:2012if}. Nevertheless, for
the computation of the energy, the equation of motion contracted
with the solution itself was simply assumed to be satisfied. To
test the validity of this assumption, we need to explicitly show
that
\begin{eqnarray}
\label{equationmotion1} \langle \Psi Q \Psi \rangle + \langle \Psi
\Psi \Psi \rangle=0.
\end{eqnarray}
In the previous paper \cite{AldoArroyo:2012if}, only the kinetic
term $\langle \Psi Q \Psi \rangle$ was computed. And therefore, it
remains the computation of the cubic term $\langle \Psi \Psi \Psi
\rangle$. This calculation will be performed in the next section.

\section{Evaluation of the cubic term of the action}

Although the solution (\ref{solpsi2}) can be written as a pure
gauge form $\Psi=U Q U^{-1}$ such that formally satisfies the string field equation of motion $Q
\Psi + \Psi \Psi =0$, it is not a trivial task to test if the equation of motion
contracted with the solution itself is satisfied. In general, a priori there is no justification for assuming
the validity of $\langle \Psi Q \Psi \rangle + \langle \Psi
\Psi \Psi \rangle=0$ without an explicit calculation. Therefore the cubic term of the action must be evaluated.

Before computing the cubic term of the action for the multi-brane
solutions. We are going to calculate a correlator that will be
very useful for the evaluation of the cubic term. The definition
of the considered correlator is as follows
\begin{eqnarray}
\label{ccc1} \Big\langle G_1,G_2,G_3\Big\rangle = \langle\langle
B G_1(K) c G_2(K) c G_3(K)\gamma^2\rangle\rangle,
\end{eqnarray}
for a general set of functions $G_i(K)$. The inclusion of notation
$\langle\langle \,\cdots \rangle\rangle$ refers for a standard
correlator with the difference that we have to insert the operator
$Y_{-2}$ at the open string midpoint. The operator $Y_{-2}$ can be
given as the product of two inverse picture changing operators,
$Y_{-2} = Y (i)Y (-i)$, with $Y (z) = - \partial \xi
e^{-2\phi}c(z)$.

Let us define all functions $G_i(K)$ as an integral representation
of a continuous superposition of wedge states,
\begin{eqnarray}
\label{fi1} G_i(K)=\int_{0}^{\infty} dt_i g_i(t_i) e^{-t_i K}.
\end{eqnarray}
Formally equation (\ref{fi1}) can be thought as a Laplace
transform. The validity of this representation depends on specific
holomorphicity conditions imposed on the functions $G_i(K)$.
Detailed discussions regarding to these conditions were studied in
reference \cite{Schnabl:2010tb}. However at this point, let us
simply assume that the functions $G_i(K)$ satisfied the preceding
conditions.

Replacing the integral representation of the functions $G_i$'s
(\ref{fi1}) into (\ref{ccc1}), we obtain the following triple
integral
\begin{align}
\label{rr1} \Big\langle G_1,G_2,G_3\Big\rangle = \int_{0}^{\infty}
dt_1 dt_2 dt_3  g_1(t_1) g_2(t_2) g_3(t_3) \langle\langle B
e^{-t_1 K}ce^{-t_2 K}ce^{-t_3 K}\gamma^2 \rangle\rangle.
\end{align}
The correlator $\langle\langle B e^{-t_1 K}ce^{-t_2 K}ce^{-t_3
K}\gamma^2 \rangle\rangle$ has been evaluated in references
\cite{Erler:2007xt,Gorbachev:2010zz,Arroyo:2010fq}
\begin{eqnarray}
\label{rr2} \langle\langle B e^{-t_1 K}ce^{-t_2 K}ce^{-t_3
K}\gamma^2 \rangle\rangle = \frac{s}{2 \pi^2} t_2, \;\;
\text{where} \;\; s= t_1+t_2+t_3.
\end{eqnarray}

Next we are going to use the $s$-$z$ trick developed in
\cite{Murata:2011ex,Murata:2011ep}. Essentially the trick tells us
to insert the identity
\begin{eqnarray}
\label{ident1} 1=\int_{0}^{\infty} ds
\delta\Big(s-\sum_{i=1}^{3}t_i\Big) =\int_{0}^{\infty}
ds\int_{-i\infty}^{+i\infty} \frac{dz}{2 \pi i} e^{sz}
e^{-z\sum_{i=1}^{3}t_i},
\end{eqnarray}
into the triple integral (\ref{rr1}). This identity allows us to
treat the variable $s$ as independent of the other integration
variables $t_i$. Employing the correlator (\ref{rr2}) and
inserting the identity (\ref{ident1}) into (\ref{rr1}), we get
\begin{align}
\label{rr3} \frac{1}{2 \pi^2}\int_{0}^{\infty} dt_1 dt_2 dt_3
g_1(t_1) t_2 g_2(t_2) g_3(t_3) \int_{0}^{\infty}
ds\int_{-i\infty}^{+i\infty} \frac{dz}{2 \pi i} s \, e^{sz}
e^{-z\sum_{i=1}^{3}t_i}.
\end{align}
Carrying out the integral over the variables $t_i$ and rewriting
the result in terms of the functions $G_i(z)$, we obtain
\begin{align}
\label{rr4}\Big\langle G_1,G_2,G_3\Big\rangle = -\frac{1}{2 \pi^2}
\int_{0}^{\infty} ds\int_{-i\infty}^{+i\infty} \frac{dz}{2 \pi i}
s \, e^{sz} G'_2(z)G_1(z)G_3(z).
\end{align}

Note that this correlator is simpler than the one derived in the
bosonic case, where trigonometric functions are involved and
produce lengthy results for the corresponding correlator
\cite{Murata:2011ex,Murata:2011ep}. With the aid of the above
formula (\ref{rr4}), we are in position to evaluate the cubic term
of the action for the multi-brane solutions.

Plugging the solution (\ref{solpsi2}) into the cubic term of the
action $\langle \Psi  \Psi \Psi \rangle$ and employing the
relations (\ref{sec208}), after performing some algebraic
manipulations, we obtain
\begin{eqnarray}
\label{cubicvvv1} \langle \Psi  \Psi \Psi \rangle = 3 \, \Big\langle
\frac{K}{G}(1-G),(1-G),\frac{K}{G}(1-G)\Big\rangle,
\end{eqnarray}
where $G= 1-F^2$.

For the correlator given on the right-hand side of equation
(\ref{cubicvvv1}) the functions $G_i$'s are identified by
$G_1=\frac{K}{G}(1-G)$, $G_2=(1-G)$ and $G_3=\frac{K}{G}(1-G)$.
Once this identification has been made, the next step is to use
the result (\ref{rr4}). And hence we arrive at the following
expression for the cubic term
\begin{eqnarray}
\label{cubicvvv2} \langle \Psi  \Psi \Psi \rangle =-\frac{1}{2
\pi^2} \int_{0}^{\infty} ds\int_{-i\infty}^{+i\infty} \frac{dz}{2
\pi i} s \, e^{sz} z^2 \Big[ \frac{6 G'(z)}{G(z)}-\frac{3
G'(z)}{G(z)^2}-3 G'(z) \Big].
\end{eqnarray}
Since the term inside the brackets does not depend on the variable
$s$, we can evaluate the integral over this variable, which is
well defined for values of the variable $z$ such that Re$(z)<0$.
Performing the integral over $s$, we obtain
\begin{eqnarray}
\label{rr5} \langle \Psi  \Psi \Psi \rangle = -\frac{1}{2 \pi^2}
\int_{-i\infty}^{+i\infty} \frac{dz}{2 \pi i} \Big[ \frac{6
G'(z)}{G(z)}-\frac{3 G'(z)}{G(z)^2}-3 G'(z) \Big].
\end{eqnarray}

At this stage, we are going to impose specific conditions on the
corresponding functions.  The motivation for demanding these
conditions, as we are going to see, will be the fact that the
energy computed from the action and from the Ellwood's gauge
invariant will agree provided that the function that parameterizes
the multi-brane solutions satisfies holomorphicity conditions that
are similar to the bosonic case \cite{Murata:2011ep}.

Let us assume that the function appearing in the expression of the
cubic term of the action (\ref{rr5}), can be written as
$G(z)=1+\sum_{n=1}^{\infty} a_n z^{-n}$, namely $G$ is holomorphic
at the point at infinity $z=\infty$ and has a limit $G(\infty)=1$.
Under this condition, it is possible to make the integral along
the imaginary axis into a sufficiently large closed contour $C$
running in the counterclockwise direction by adding a large
non-contributing half-circle in the left half plane such that
Re$(z) < 0$, and consequently the integral (\ref{rr5}) can be
written as
\begin{eqnarray}
\label{rr6} \langle \Psi  \Psi \Psi \rangle = -\frac{1}{2 \pi^2}
\oint_{C} \frac{dz}{2 \pi i} \Big[ \frac{6 G'(z)}{G(z)}-\frac{3
G'(z)}{G(z)^2}-3 G'(z) \Big].
\end{eqnarray}

Moreover by demanding two additional requirements for the
functions $G$ and $1/G$,
\begin{itemize}
    \item $G$ and $1/G$ are holomorphic in Re$(z) \geq 0$ except at
    $z=0$.
    \item $G$ or $1/G$ are meromorphic at $z=0$.
   \end{itemize}
We can stretch the $C$ contour around infinity, picking up only a
possible contribution from the origin,
\begin{eqnarray}
\label{cubic1x5} \langle \Psi  \Psi \Psi \rangle = -\frac{1}{2
\pi^2} \oint_{C_0} \frac{dz}{2 \pi i} \Big[ \frac{6 G'(z)}{G(z)}+3
\partial_z \big\{ \frac{1}{G(z)}-G(z) \big\} \Big],
\end{eqnarray}
where $C_0$ is a contour encircling the origin in the clockwise
direction. As shown explicitly, the second term appearing in the
integrand on the right-hand side of (\ref{cubic1x5}) is a total
derivative term with respect to $z$ such that the contour integral
of that term usually vanishes. In fact, since we assume the
meromorphicity of $G(z)$ at the origin, this total derivative term
vanishes. Now inverting the direction of the contour $C_0$, we
finally obtain
\begin{eqnarray}
\label{cubic1x6} \langle \Psi  \Psi \Psi\rangle = \frac{3}{ \pi^2}
\oint \frac{dz}{2 \pi i} \frac{ G'(z)}{G(z)} .
\end{eqnarray}
In order to calculate the contour integral (\ref{cubic1x6}), we
need to follow a closed curve encircling the origin in the
counterclockwise direction.

Let us remember that under the same holomorphicity conditions
satisfied by the function that parameterizes the multi-brane
solutions, the kinetic term of the action was computed in
reference \cite{AldoArroyo:2012if}
\begin{eqnarray}
\label{kine1x6} \langle \Psi Q\Psi \rangle = -\frac{3}{ \pi^2}
\oint \frac{dz}{2 \pi i} \frac{ G'(z)}{G(z)} .
\end{eqnarray}
Therefore adding equations (\ref{cubic1x6}) and (\ref{kine1x6}),
we conclude that the assumption of the validity of the equation of
motion contracted with the solution itself was correct provided
that the function that parameterizes the multi-brane solutions
satisfies the aforementioned holomorphicity requirements.

\subsection{Discussing the result for the cubic term}
The final result (\ref{cubic1x6}) for the cubic term relies on the
validity of the step from equation (\ref{cubicvvv2}) to
(\ref{rr5}). The integrand in equation (\ref{cubicvvv2}) can have
poles at $z=0$ for a function $G(z)$ satisfying the three
holomorphicity conditions previously given. To avoid the
singularities at $z=0$, we have simply shifted the integration
over $z$, which is originally along Re$(z)=0$, to that along
Re$(z)<0$. This procedure needs to be justified.

A similar observation for the result in the bosonic case
\cite{Murata:2011ex,Murata:2011ep} has been made in Hata and
Kojita's paper \cite{Hata:2012cy}. To treat the points at $z=0$,
we use the property that the eigenvalue distribution of $K$ is
restricted to real and non-negative
\cite{Hata:2011ke,Hata:2012cy}, and so we can replace $K
\rightarrow K+\epsilon$, with $\epsilon$ being a positive
infinitesimal. Now if we compute the cubic term with $K$ replaced
by $K+\epsilon$ and take the limit $\epsilon \rightarrow 0$, we
obtain
\begin{eqnarray}
\label{cubicvvvepsilon0} \langle \Psi  \Psi \Psi \rangle
=-\frac{1}{2 \pi^2} \int_{0}^{\infty} ds\int_{-i\infty}^{+i\infty}
\frac{dz}{2 \pi i} s \, e^{sz} z^2 \Big[ \frac{6
G'(z)}{G(z)}-\frac{3 G'(z)}{G(z)^2}-3 G'(z) \Big],
\end{eqnarray}
where the integration over $z$ is along a line parallel to the
pure-imaginary axis with Re$(z)>0$. Since $\epsilon >0$, it is
easy to see why in this case the integration must be along
Re$(z)>0$.

In order to simplify the notation, let us define the function
$J(z)$ as
\begin{eqnarray}
\label{Jz1} J(z)= -\frac{z^2}{2 \pi^2}  \Big[ \frac{6
G'(z)}{G(z)}-\frac{3 G'(z)}{G(z)^2}-3 G'(z) \Big].
\end{eqnarray}
Employing this definition (\ref{Jz1}) into equation
(\ref{cubicvvvepsilon0}), we write the cubic term as follows
\begin{eqnarray}
\label{cubiccontor1} \langle \Psi  \Psi \Psi \rangle =
\int_{0}^{\infty} ds\int_{\mathcal{C}_{>}} \frac{dz}{2 \pi i} s \,
e^{sz} J(z),
\end{eqnarray}
where the notation $\mathcal{C}_{>}$ represents the curve
corresponding to the line parallel to the pure-imaginary axis with
Re$(z)>0$. Let us also denote $\mathcal{C}_{<}$ as the curve
corresponding to the line parallel to the pure-imaginary axis with
Re$(z)<0$. Note that the integration over $z$ along the curve
$\mathcal{C}_{<}$ corresponds to the one used in passing of the
step from equation (\ref{cubicvvv2}) to (\ref{rr5}).

By inverting the direction of the curve $\mathcal{C}_{<}$ and
joining its end points with the end points of curve
$\mathcal{C}_{>}$, we construct a large closed curve running in
the counterclockwise direction. Since the integrand $s e^{s z}
J(z)$ can have poles at $z=0$, the integration over $z$ along this
large closed curve is equivalent to the integration along a closed
curve encircling the origin in the counterclockwise direction,
\begin{eqnarray}
\label{cubiccontor22} \int_{0}^{\infty} ds\int_{\mathcal{C}_{>}}
\frac{dz}{2 \pi i} s \, e^{sz} J(z) - \int_{0}^{\infty}
ds\int_{\mathcal{C}_{<}} \frac{dz}{2 \pi i} s \, e^{sz} J(z) =
\int_{0}^{\infty} ds\oint \frac{dz}{2 \pi i} s \, e^{sz} J(z) ,
\end{eqnarray}
where we have the minus sign because by construction the left hand
side of the large closed curve goes in the opposite direction of
$\mathcal{C}_{<}$.

Employing equations (\ref{cubiccontor1}) and
(\ref{cubiccontor22}), we see that the cubic term of the action is
given by
\begin{eqnarray}
\label{cubiccontor33} \langle \Psi  \Psi \Psi \rangle =
\int_{0}^{\infty} ds\int_{\mathcal{C}_{<}} \frac{dz}{2 \pi i} s \,
e^{sz} J(z) + \int_{0}^{\infty} ds\oint \frac{dz}{2 \pi i} s \,
e^{sz} J(z) .
\end{eqnarray}
The first term on the right hand side of equation
(\ref{cubiccontor33}) precisely corresponds to the term on the
right hand side of equation (\ref{cubicvvv2}), with the
integration over $z$ along Re$(z)<0$. The desired result
(\ref{rr5}) is obtained provided that the second term on the right
hand side of equation (\ref{cubiccontor33}) vanishes. Thus, we
need to prove that $\mathcal{I}=0$, where $\mathcal{I}$ is defined
as the following integral
\begin{eqnarray}
\label{zero11} \mathcal{I} = \int_{0}^{\infty} ds\oint \frac{dz}{2
\pi i} s \, e^{sz} J(z).
\end{eqnarray}

The result given by equation (\ref{cubiccontor33}) is quite
similar to the one obtained in the bosonic context
\cite{Hata:2012cy}. Actually, using the
$K_{\epsilon}$-regularization and the function
\begin{eqnarray}
\label{GGKK} G(K)=\Big( \frac{K}{1+K} \Big)^n,
\end{eqnarray}
the evaluation of the cubic term leads to the result
\begin{eqnarray}
\label{sec301} \frac{\pi^2}{3} \langle \Psi  \Psi \Psi
\rangle_{\text{bosonic}} = n + \mathcal{A}_n,
\end{eqnarray}
with
\begin{eqnarray}
\label{sec106} \mathcal{A}_n= \frac{\pi^2}{3} n (1-n^2)
\text{Re\,} _1\mathcal{F}_1(2-n,4;2 \pi i) ,
\end{eqnarray}
where $_1\mathcal{F}_1(a,b; z)$ is the confluent hypergeometric
function. Note that the expected result, $\frac{\pi^2}{3} \langle
\Psi \Psi \Psi \rangle_{\text{bosonic}} = n$, is obtained only for
values of $n$ such that $n=0,\pm 1$. Let us see what happens for
the superstring case.

Performing the replacement $K \rightarrow K+\epsilon$, we obtain
the following expression for the integral $ \mathcal{I}
\rightarrow \mathcal{I}_\epsilon $
\begin{eqnarray}
\label{zero11epsi1} \mathcal{I}_\epsilon = \int_{0}^{\infty}
ds\oint \frac{dz}{2 \pi i} s \, e^{(z-\epsilon) s} J(z),
\end{eqnarray}
where the part $e^{-\epsilon s}$ comes from $K \rightarrow
K+\epsilon$. To evaluate this integral (\ref{zero11epsi1}), we are
going to use the function (\ref{GGKK}) which satisfies the
aforementioned three holomorphicity conditions. This is the same
function that has been used in the analysis of multibrane
solutions in the bosonic case
\cite{Murata:2011ex,Murata:2011ep,Hata:2011ke,Hata:2012cy}.

Since the $z$-integration is a contour integral performed around a
closed curve encircling the origin in the counterclockwise
direction, to compute the integral over this variable $z$, we need
to write the Laurent series of the integrand around $z=0$
\begin{eqnarray}
\label{expandgn1} s \, e^{(z-\epsilon) s} J(z) =
\frac{I_n(s,\epsilon)}{z} + \sum_{p \neq -1} I_{p,n}(s,\epsilon)
z^p,
\end{eqnarray}
and pick up the coefficient $I_n(s,\epsilon)$ in front of the term
$1/z$. Then by performing the $s$-integration, we obtain the value
of the integral (\ref{zero11epsi1}), namely $\mathcal{I}_\epsilon
= \int_{0}^{\infty} ds\, I_n(s,\epsilon)$.

With the aid of equations (\ref{Jz1}), (\ref{GGKK}) and
(\ref{expandgn1}), we are in position to explicitly evaluate the
coefficient $I_n(s,\epsilon)$ for various values of $n$. It turns
out that the coefficient $I_n(s,\epsilon)$ vanishes identically
for values of $n=0,\pm 1$. Let us see what happens if $|n| \geq
2$. For instance, with the values of $n=\pm 2, \pm 3,\pm 4$, we
obtain the following expressions for the coefficients
\begin{eqnarray}
\label{C_2} I_{\pm2}(s,\epsilon) &=& \pm \frac{3 s e^{-s \epsilon }}{\pi ^2}, \\
\label{C_3} I_{\pm3}(s,\epsilon) &=& \pm \frac{9 s (s+2) e^{-s \epsilon }}{2 \pi ^2}, \\
\label{C_4} I_{\pm4}(s,\epsilon) &=& \pm \frac{3 s \left(s^2+6
s+6\right) e^{-s \epsilon }}{\pi ^2}.
\end{eqnarray}
Since $\mathcal{I}_\epsilon = \int_{0}^{\infty} ds\,
I_n(s,\epsilon)$, the value of the integral $ \mathcal{I}_\epsilon
\rightarrow \mathcal{I} $ in the limit $\epsilon \rightarrow 0$ is
\textit{non-vanishing and divergent} except for the cases where
$n=0,\pm 1$. In fact, with $n=\pm 2$ we obtain
$\mathcal{I}_\epsilon = \pm (3/\pi^2)\int_{0}^{\infty} ds\, s
e^{-s \epsilon }  \propto 1/\epsilon^2$. This result, together
with equation (\ref{cubiccontor33}) implies that the validity of
the step from equation (\ref{cubicvvv2}) to (\ref{rr5}) only
follows when $n=0,\pm 1$.

\section{Evaluation of the Ellwood's gauge invariant}

In this section, the Ellwood's gauge invariant overlap for the
multi-brane solutions will be evaluated. A similar computation was
done in reference \cite{Erler:2010pr} for the half-brane solution.
The Ellwood's gauge invariant overlap is given by
\begin{eqnarray}
\label{inv01} W(\Psi,\mathcal{V}) = \text{Tr}(\Psi),
\end{eqnarray}
where the notation Tr$(\cdots)$ is defined in the same way as the
correlator (\ref{ccc1}) except the picture changing operator
$Y_{- 2}$ is replaced by an on shell closed string vertex operator
$\mathcal{V}( i )$ inserted at the midpoint,
$\text{Tr}(\Psi)=\langle \mathcal{V}( i ) \Psi \rangle$. We assume
the same $\mathcal{V}$ used in reference \cite{Erler:2010pr}, this
field is an NS-NS closed string vertex operator of the form
\begin{eqnarray}
\label{inv02} \mathcal{V}(z)= c \tilde{c} e^{-\phi}
e^{-\tilde{\phi}} \mathcal{O}^{\text{m}},
\end{eqnarray}
where $\mathcal{O}^{\text{m}}$ is a weight $(\frac{1}{2},
\frac{1}{2})$ superconformal matter primary field. As argued by
Ellwood \cite{Ellwood:2008jh}, the gauge invariant overlap
represents the shift in the closed string tadpole of the solution
relative to the perturbative vacuum.

Replacing the multi-brane solution (\ref{solpsi2}) into the
definition of the gauge invariant overlap (\ref{inv01}), the term
$\text{Tr}(F B \gamma^2 F)$ does not contribute since we need
three $c$'s fields to saturate the corresponding correlator, and
the insertion $\mathcal{V}$ already has two $c$'s fields.
Therefore, we obtain
\begin{eqnarray}
\label{inv03} W(\Psi,\mathcal{V}) = \text{Tr}(F c \frac{KB}{1-F^2}
c F).
\end{eqnarray}
As for the evaluation of the cubic term of the action, let us
write the functions $F$ and $K/G$ as an integral representation of
a continuous superposition of wedge states,
\begin{eqnarray}
\label{inv04} F&=&\int_{0}^{\infty} dt f(t)
e^{-t K}, \\
\label{inv05} K/G&=&\int_{0}^{\infty} dt g(t) e^{-t K},
\end{eqnarray}
where $G=1-F^2$. The validity of this assumption depends on the
holomorphicity conditions satisfied by the functions. Replacing
equations (\ref{inv04}) and (\ref{inv05}) into (\ref{inv03}), we
obtain
\begin{eqnarray}
\label{inv07} W(\Psi,\mathcal{V}) = \int_{0}^{\infty} dt_1
 dt_2 dt_3 f(t_1) g(t_2) f(t_3) \text{Tr}(
e^{-t_1 K} c e^{-t_2 K} B c e^{-t_3 K}).
\end{eqnarray}
The correlator $\text{Tr}( e^{-t_1 K} c e^{-t_2 K} B c e^{-t_3
K})$ has been evaluated in reference \cite{Erler:2010pr} by using
the usual scaling argument \cite{Ellwood:2008jh}
\begin{eqnarray}
\label{inv08} \text{Tr}( e^{-t_1 K} c e^{-t_2 K} B c e^{-t_3 K}) =
(t_1+t_3) \text{Tr}( c \Omega) ,
\end{eqnarray}
where $\Omega = e^{-K}$ and $\text{Tr}( c \Omega) = \langle
\mathcal{V}(i \infty) c(0)\rangle_{C_{1}}$ is the expected result
of the closed string tadpole on the disk.

Replacing the correlator (\ref{inv08}) into (\ref{inv07}), we
obtain the following expression for the gauge invariant overlap
\begin{eqnarray}
\label{inv09} W(\Psi,\mathcal{V}) = \int_{0}^{\infty} dt_1
 dta_2 dt_3 f(t_1) g(t_2) f(t_3) (t_1+t_3) \text{Tr}( c \Omega).
\end{eqnarray}

To evaluate the right-hand side of equation (\ref{inv09}), we use
again the $s$-$z$ trick. Inserting the identity (\ref{ident1})
into the triple integral (\ref{inv09}), we obtain
\begin{align}
\label{corre1f2x1xx} \int_{0}^{\infty} dt_1 dt_2 dt_3 f(t_1)
(t_1+t_3) g(t_2) f(t_3) \int_{0}^{\infty}
ds\int_{-i\infty}^{+i\infty} \frac{dz}{2 \pi i} \, e^{sz}
e^{-z\sum_{i=1}^{3}t_i}\text{Tr}( c \Omega) .
\end{align}
Evaluating the integral over the variables $t_i$ and rewriting the
result in terms of the functions $F(z)$ and $z/G(z)$, we get
\begin{align}
\label{corre1f2x2xx} W(\Psi,\mathcal{V}) &= - 2\int_{0}^{\infty}
ds\int_{-i\infty}^{+i\infty} \frac{dz}{2 \pi i} \, e^{sz} z \frac{
F(z) F'(z)}{G(z)}\text{Tr}( c \Omega) \nonumber \\
&= \int_{0}^{\infty} ds\int_{-i\infty}^{+i\infty} \frac{dz}{2 \pi
i} \, e^{sz} z \frac{ G'(z)}{G(z)}\text{Tr}( c \Omega),
\end{align}
where $G(z)=1-F^2(z)$. Evaluating the integral over the variable
$s$, which is well defined for Re$(z)<0$, we obtain
\begin{eqnarray}
\label{tadpolexx1} W(\Psi,\mathcal{V}) = -
\int_{-i\infty}^{+i\infty} \frac{dz}{2 \pi i} \frac{G'(z)}{G(z)}
\text{Tr}( c \Omega).
\end{eqnarray}

Employing the same holomorphicity conditions used in the
evaluation of the cubic term of the action, we can take the
integral along the imaginary axis into a sufficiently large closed
contour $C$ running in the counterclockwise direction by adding a
large non-contributing half-circle in the left half plane Re$(z) <
0$. So that the Ellwood's gauge invariant overlap for the
multi-brane solution (\ref{solpsi2}) can be written as the
following contour integral
\begin{eqnarray}
\label{tadpolexx2}  W(\Psi,\mathcal{V})  = - \oint_{C} \frac{dz}{2
\pi i} \frac{G'(z)}{G(z)} \text{Tr}( c \Omega).
\end{eqnarray}

Furthermore we can stretch the $C$ contour around infinity,
picking up only a possible contribution from the origin,
\begin{eqnarray}
\label{tadpolex5} W(\Psi,\mathcal{V})  = - \oint_{C_0} \frac{dz}{2
\pi i} \frac{G'(z)}{G(z)} \text{Tr}( c \Omega),
\end{eqnarray}
where $C_0$ is a contour encircling the origin in the clockwise
direction. Now inverting the direction of the contour $C_0$, we
finally obtain
\begin{eqnarray}
\label{tadpolex6} W(\Psi,\mathcal{V})  =  \oint \frac{dz}{2 \pi i}
\frac{G'(z)}{G(z)} \text{Tr}( c \Omega) .
\end{eqnarray}
As in the case of the expression for the cubic term
(\ref{cubic1x6}), to compute the contour integral
(\ref{tadpolex6}), we need to follow a closed curve encircling the
origin in the counterclockwise direction.

Note that the final result for the Ellwood's gauge invariant
(\ref{tadpolex6}) depends on the holomorphicity conditions imposed
on the function that parameterizes the multi-brane solutions. As
in the bosonic case, it should be nice to analyze if the violation
of some of these holomorphicity conditions leads to the appearance
of anomalies associated to the evaluation of the gauge-invariant
observable \cite{Murata:2011ep,Hata:2011ke,Hata:2012cy}.

\subsection{Discussing the result for the Ellwood's gauge invariant}
The final result for the Ellwood's gauge invariant
(\ref{tadpolex6}) relies on the validity of the step from equation
(\ref{corre1f2x2xx}) to (\ref{tadpolexx1}). The integrand in
equation (\ref{corre1f2x2xx}) can have poles at $z=0$ for a
function $G(z)$ satisfying the three holomorphicity conditions
previously given. To avoid the singularities at $z=0$, we have
simply shifted the integration over $z$, which is originally along
Re$(z)=0$, to that along Re$(z)<0$. As in the case of the cubic
term, we need to justify this procedure.

Employing the same arguments developed for the case of the cubic
term, we show that the Ellwood's gauge invariant can be written as
\begin{eqnarray}
\label{tadpolezzy1} W(\Psi,\mathcal{V}) = \int_{0}^{\infty}
ds\int_{\mathcal{C}_{<}} \frac{dz}{2 \pi i}  \, e^{sz} z
\frac{G'(z)}{G(z)} \text{Tr}( c \Omega) + \int_{0}^{\infty}
ds\oint \frac{dz}{2 \pi i} \, e^{sz} z \frac{G'(z)}{G(z)}
\text{Tr}( c \Omega) .
\end{eqnarray}
The first term on the right hand side of equation
(\ref{tadpolezzy1}) precisely corresponds to the term on the right
hand side of equation (\ref{corre1f2x2xx}), with the integration
over $z$ along Re$(z)<0$. The desire result (\ref{tadpolexx1}) is
obtained provided that the second term on the right hand side of
equation (\ref{tadpolezzy1}) vanishes. Therefore, we need to prove
that $\mathcal{K}=0$, where $\mathcal{K}$ is defined by
\begin{eqnarray}
\label{zeroEll1} \mathcal{K} = \int_{0}^{\infty} ds\oint
\frac{dz}{2 \pi i} \, e^{sz} z \frac{G'(z)}{G(z)} \text{Tr}( c
\Omega).
\end{eqnarray}

As for the cubic term, by performing the replacement $K
\rightarrow K+\epsilon$, we obtain the following expression for
the integral $ \mathcal{K} \rightarrow \mathcal{K}_\epsilon $
\begin{eqnarray}
\label{zeroEll2y} \mathcal{K}_{\epsilon} = \int_{0}^{\infty}
ds\oint \frac{dz}{2 \pi i} \, e^{s(z-\epsilon)} z
\frac{G'(z)}{G(z)} \text{Tr}( c \Omega).
\end{eqnarray}
where the part $e^{-\epsilon s}$ comes from $K \rightarrow
K+\epsilon$. To evaluate this integral (\ref{zeroEll2y}), let us
use the function $G(z)$ defined by equation (\ref{GGKK}).

Since the $z$-integration is a contour integral performed around a
closed curve encircling the origin in the counterclockwise
direction, to compute the integral over this variable $z$, we need
to write the Laurent series of the integrand around $z=0$
\begin{eqnarray}
\label{expandEllcoef1} e^{s(z-\epsilon)} z \frac{G'(z)}{G(z)}
\text{Tr}( c \Omega) = \Big[ \frac{\mathcal{K}_n(s,\epsilon)}{z} +
\sum_{p \neq -1} \mathcal{K}_{p,n}(s,\epsilon) z^p \Big]
\text{Tr}( c \Omega),
\end{eqnarray}
and pick up the coefficient $\mathcal{K}_n(s,\epsilon)$ in front
of the term $1/z$. Then by performing the $s$-integration, we
obtain the value of the integral (\ref{zeroEll2y}), namely
$\mathcal{K}_\epsilon = \int_{0}^{\infty} ds\,
\mathcal{K}_n(s,\epsilon)\text{Tr}( c \Omega)$.

With the aid of equations (\ref{GGKK}) and (\ref{expandEllcoef1}),
we are in position to explicitly evaluate the coefficient
$\mathcal{K}_n(s,\epsilon)$ for various values of $n$. It turns
out that the coefficient $\mathcal{K}_n(s,\epsilon)$ vanishes
identically for any integer value of $n$ (while for the case of
the cubic term, only the coefficients with $n=0,\pm 1$ vanish
identically). Since $ \mathcal{K}_\epsilon \rightarrow \mathcal{K}
$ in the limit $\epsilon \rightarrow 0$, we conclude that
$\mathcal{K} = 0$. This result together with equation
(\ref{tadpolezzy1}) justify the validity of the step from equation
(\ref{corre1f2x2xx}) to (\ref{tadpolexx1}).

\section{Summary and conclusions}

Given the following list of holomorphicity conditions imposed on the function that
parameterizes the multi-brane solutions
\begin{enumerate}[i)]
    \item $G$ and $1/G$ are holomorphic in Re$(z) \geq 0$ except at
    $z=0$.
    \item $G$ or $1/G$ are meromorphic at $z=0$.
    \item $G$ is holomorphic
at the point at infinity $z=\infty$ and has a limit $G(\infty)=1$.
   \end{enumerate}
We have evaluated the cubic term of action for the multi-brane solutions. The result
is given in terms of a contour integral
\begin{eqnarray}
\label{cubic1x6final} \langle \Psi  \Psi \Psi\rangle = \frac{3}{ \pi^2}
\oint \frac{dz}{2 \pi i} \frac{ G'(z)}{G(z)} .
\end{eqnarray}

Now by employing the result coming from the evaluation of kinetic term of the action,
which has been performed in reference \cite{AldoArroyo:2012if}
\begin{eqnarray}
\label{kine1x6final} \langle \Psi Q\Psi \rangle = -\frac{3}{ \pi^2}
\oint \frac{dz}{2 \pi i} \frac{ G'(z)}{G(z)} ,
\end{eqnarray}
we can write the following expression for the energy
\begin{eqnarray}
\label{energy1f} E = \frac{1}{2} \langle \Psi Q \Psi \rangle +
\frac{1}{3} \langle \Psi \Psi \Psi \rangle = -\frac{1}{2 \pi^2}
\oint \frac{dz}{2 \pi i} \frac{G'(z)}{G(z)}  .
\end{eqnarray}

Using the same holomorphicity conditions i)-iii), we have also
computed the Ellwood's gauge invariant overlap for the multi-brane
solutions and we have found the result
\begin{eqnarray}
\label{tadpole1f} W(\Psi,\mathcal{V})  =  \oint \frac{dz}{2 \pi i}
\frac{G'(z)}{G(z)} \text{Tr}( c \Omega) .
\end{eqnarray}
Let us remember that to compute the above contour integrals, we
need to follow a closed curve encircling the origin in the
counterclockwise direction.

Comparing equations (\ref{energy1f}) and (\ref{tadpole1f}), we
conclude that the energy computed from the action and from the
Ellwood's invariant will agree provided that the function that
parameterizes the multi-brane solutions satisfies the
holomorphicity conditions i)-iii). This conclusion turns out to be
true as long as the values of the integer $n$ appearing in the
definition of the function $G(z)=[ z/(1+z) ]^n$ are restricted to
the values $n=0,\pm 1$. This result is similar to the bosonic case
\cite{Hata:2012cy}.

Prior the proposed multi-brane solutions, in the framework of the
modified cubic superstring field theory, solutions of the form
\begin{eqnarray}
\label{solmulti1} \Psi= F c \frac{KB}{1-F^2} c F + F B \gamma^2 F,
\end{eqnarray}
have been considered for the specific cases: $F^2=e^{-K}$ and
$F^2=1/(1+K)$, where it was shown that the solutions characterize
the tachyon vacuum solution \cite{Erler:2007xt,Gorbachev:2010zz}.
It is interesting to note that, as argued in reference
\cite{Erler:2007xt}, from an analytic perspective the suggested
tachyon vacuum solution appears to be as regular as Schnabl's
original solution in the open bosonic string field theory
\cite{Schnabl:2005gv}. Nevertheless, from the perspective of the
level expansion the situation is unclear, though to be honest, the
analysis of the energy for the tachyon vacuum solution using the
usual Virasoro $L_0$ level expansion has not yet been carried out.
In this respect, the situation for the multi-brane solutions is
similar and therefore it should be a good research project to
analyze the solutions using the Virasoro $L_0$ level expansion.

Finally, we would like to comment about Berkovits non-polynomial
open superstring field theory \cite{Berkovits:1995ab}, since this
theory is based on Witten's associative star product, its
mathematical setup shares the same algebraic structure of both
string field theories, the open bosonic string field theory and
the modified cubic superstring field theory, and hence the
strategy and prescriptions studied in this work should be directly
extended to that theory. Recently, the construction of the tachyon
vacuum solution in Berkovits superstring field theory based on
elements in the $KBc\gamma\gamma^{-1}$ subalgebra has been
proposed by T. Erler \cite{Erler:2013wda}.

\section*{Acknowledgements}
I would like to thank Ted Erler, Masaki Murata, Martin Schnabl and
Michael Kroyter for useful discussions. This work is supported by
CNPq grant 303073/2012-8.



\begin{thebibliography}
\bibitem{Schnabl:2005gv}
  M.~Schnabl, \textit{Analytic solution for tachyon condensation in open string field
theory},
  Adv.\ Theor.\ Math.\ Phys.\  {\bf 10}, 433 (2006), hep-th/0511286.

\bibitem{Okawa:2006vm}
  Y.~Okawa,
  \textit{Comments on Schnabl's analytic solution for tachyon condensation in
  Witten's open string field theory},
  JHEP {\bf 0604}, 055 (2006), hep-th/0603159.

\bibitem{Fuchs:2006hw}
  E.~Fuchs and M.~Kroyter, \textit{On the validity of the solution of string field theory},
  JHEP {\bf 0605}, 006 (2006),
  hep-th/0603195.

\bibitem{Rastelli:2006ap}
  L.~Rastelli and B.~Zwiebach, \textit{Solving Open String Field Theory with Special Projectors},
  JHEP {\bf 0801}, 020 (2008), hep-th/0606131.

\bibitem{Okawa:2006sn}
  Y.~Okawa, L.~Rastelli and B.~Zwiebach, \textit{Analytic Solutions for Tachyon Condensation with General
Projectors},
  hep-th/0611110.

\bibitem{Erler:2012qr}
  T.~Erler and C.~Maccaferri, \textit{The Phantom Term in Open String Field Theory},
  JHEP {\bf 1206}, 084 (2012),
  [arXiv:1201.5122].

\bibitem{Ellwood:2009zf}
  I.~Ellwood, \textit{Singular gauge transformations in string field theory},
  JHEP {\bf 0905}, 037 (2009),
  [arXiv:0903.0390].

\bibitem{Schnabl:2010tb}
  M.~Schnabl,
  \textit{Algebraic solutions in Open String Field Theory - a lightning review},
  [arXiv:1004.4858].

\bibitem{Erler:2009uj}
  T.~Erler and M.~Schnabl,
  \textit{A Simple Analytic Solution for Tachyon Condensation},
  JHEP {\bf 0910}, 066 (2009),
  [arXiv:0906.0979].

\bibitem{AldoArroyo:2009hf}
  E.~Aldo Arroyo,
  \textit{The Tachyon Potential in the Sliver Frame},
  JHEP {\bf 0910}, 056 (2009),
  [arXiv:0907.4939].

\bibitem{Arroyo:2009ec}
  E.~A.~Arroyo,
  \textit{Cubic interaction term for Schnabl's solution using Pade approximants},
  J.\ Phys.\ A  {\bf 42}, 375402 (2009),
  [arXiv:0905.2014].

\bibitem{Arroyo:2011zt}
  E.~A.~Arroyo,
  \textit{Conservation laws and tachyon potentials in the sliver frame},
  JHEP {\bf 1106}, 033 (2011),
  [arXiv:1103.4830].

\bibitem{Bonora:2013cya}
  L.~Bonora and S.~Giaccari, \textit{Generalized states in SFT},
  [arXiv:1304.2159].

\bibitem{Masuda:2012kt}
  T.~Masuda, T.~Noumi and D.~Takahashi, \textit{Constraints on a class of classical solutions in open string field
theory}, JHEP {\bf 1210}, 113 (2012),
  [arXiv:1207.6220].

\bibitem{Erler:2006hw}
  T.~Erler,
  \textit{Split string formalism and the closed string vacuum},
  JHEP {\bf 0705}, 083 (2007), hep-th/0611200.

\bibitem{Erler:2006ww}
  T.~Erler,
  \textit{Split string formalism and the closed string vacuum. II},
  JHEP {\bf 0705}, 084 (2007), hep-th/0612050.

\bibitem{Zeze:2011zz}
  S.~Zeze,
  \textit{Application of KBc subalgebra in string field theory},
  Prog.\ Theor.\ Phys.\ Suppl.\  {\bf 188}, 56 (2011).

\bibitem{Zeze:2010jv}
  S.~Zeze,
  \textit{Tachyon potential in KBc subalgebra},
  Prog.\ Theor.\ Phys.\  {\bf 124}, 567 (2010),
  [arXiv:1004.4351].

\bibitem{Arroyo:2010sy}
  E.~A.~Arroyo,
  \textit{Comments on regularization of identity based solutions in string field
  theory},
  JHEP {\bf 1011}, 135 (2010),
  [arXiv:1009.0198].

\bibitem{AldoArroyo:2011gx}
  E.~Aldo Arroyo,
  \textit{Level truncation analysis of regularized identity based solutions},
  JHEP {\bf 1111}, 079 (2011),
  [arXiv:1109.5354].

\bibitem{Murata:2011ex}
  M.~Murata and M.~Schnabl,
  \textit{On Multibrane Solutions in Open String Field Theory},
  Prog.\ Theor.\ Phys.\ Suppl.\  {\bf 188}, 50 (2011),
  [arXiv:1103.1382].

\bibitem{Murata:2011ep}
  M.~Murata and M.~Schnabl, \textit{Multibrane Solutions in Open String Field Theory},
  JHEP {\bf 1207}, 063 (2012),
  [arXiv:1112.0591].

\bibitem{Hata:2011ke}
  H.~Hata and T.~Kojita,
  \textit{Winding Number in String Field Theory},
  JHEP {\bf 1201}, 088 (2012),
  [arXiv:1111.2389].

\bibitem{Hata:2012cy}
  H.~Hata and T.~Kojita,
  \textit{Singularities in K-space and Multi-brane Solutions in Cubic String Field Theory},
  JHEP {\bf 1302}, 065 (2013),
  [arXiv:1209.4406].

\bibitem{Arefeva:1989cp}
  I.~Y.~Arefeva, P.~B.~Medvedev and A.~P.~Zubarev,
  \textit{New Representation For String Field Solves The Consistency Problem For Open
  Superstring Field Theory},
  Nucl.\ Phys.\  B {\bf 341}, 464 (1990).

\bibitem{Witten:1985cc}
  E.~Witten,
  \textit{Noncommutative Geometry And String Field Theory},
  Nucl.\ Phys.\  B {\bf 268}, 253 (1986).

\bibitem{Erler:2007xt}
  T.~Erler,
  \textit{Tachyon Vacuum in Cubic Superstring Field Theory},
  JHEP {\bf 0801}, 013 (2008),
  [arXiv:0707.4591].

\bibitem{Kroyter:2009bg}
  M.~Kroyter,
  \textit{Comments on superstring field theory and its vacuum solution},
  JHEP {\bf 0908}, 048 (2009),
  [arXiv:0905.3501].


\bibitem{Aref'eva:2008ad}
  I.~Y.~Aref'eva, R.~V.~Gorbachev and P.~B.~Medvedev,
  \textit{Tachyon Solution in Cubic Neveu-Schwarz String Field Theory},
  Theor.\ Math.\ Phys.\  {\bf 158}, 320 (2009),
  [arXiv:0804.2017].

\bibitem{Gorbachev:2010zz}
  R.~V.~Gorbachev,
  \textit{New solution of the superstring equation of motion},
  Theor.\ Math.\ Phys.\  {\bf 162}, 90 (2010),
  [Teor.\ Mat.\ Fiz.\  {\bf 162}, 106 (2010)].

\bibitem{Erler:2010pr}
  T.~Erler, \textit{Exotic Universal Solutions in Cubic Superstring Field Theory},
  JHEP {\bf 1104}, 107 (2011), [arXiv:1009.1865].

\bibitem{Aref'eva:2009sj}
  I.~Y.~Aref'eva, R.~V.~Gorbachev and P.~B.~Medvedev,
  \textit{Pure Gauge Configurations and Solutions to Fermionic Superstring Field
  Theories Equations of Motion},
  J.\ Phys.\ A  {\bf 42}, 304001 (2009),
  [arXiv:0903.1273].

\bibitem{Arefeva:2010yd}
  I.~Y.~Arefeva and R.~V.~Gorbachev,
  \textit{On Gauge Equivalence of Tachyon Solutions in Cubic Neveu-Schwarz String
  Field Theory},
  Theor.\ Math.\ Phys.\  {\bf 165}, 1512 (2010),
  [arXiv:1004.5064].

\bibitem{Arroyo:2010fq}
  E.~A.~Arroyo,
  \textit{Generating Erler-Schnabl-type Solution for Tachyon Vacuum in Cubic
  Superstring Field Theory},
  J.\ Phys.\ A  {\bf 43}, 445403 (2010),
  [arXiv:1004.3030].

\bibitem{Inatomi:2011an}
  S.~Inatomi, I.~Kishimoto and T.~Takahashi, \textit{Homotopy Operators and Identity-Based Solutions in Cubic
Superstring Field Theory},
  JHEP {\bf 1110}, 114 (2011),
  [arXiv:1109.2406].

\bibitem{AldoArroyo:2012if}
  E.~Aldo Arroyo, \textit{Multibrane solutions in cubic superstring field theory},
  JHEP {\bf 1206}, 157 (2012),
  [arXiv:1204.0213].

\bibitem{Ellwood:2008jh}
  I.~Ellwood, \textit{The Closed string tadpole in open string field theory},
  JHEP {\bf 0808}, 063 (2008), [arXiv:0804.1131].

\bibitem{Masuda:2012cj}
  T.~Masuda, \textit{Comments on new multiple-brane solutions based on Hata-Kojita
duality in open string field theory},
  [arXiv:1211.2649].

\bibitem{Berkovits:1995ab}
  N.~Berkovits,
  \textit{SuperPoincare invariant superstring field theory},
  Nucl.\ Phys.\  B {\bf 450}, 90 (1995),
  [Erratum-ibid.\  B {\bf 459}, 439 (1996)], hep-th/9503099.

\bibitem{Erler:2013wda}
  T.~Erler, \textit{Analytic Solution for Tachyon Condensation in Berkovits' Open
Superstring Field Theory}, [arXiv:1308.4400].

\end{thebibliography}
\end{document}